\documentclass[aps, prd, superscriptaddress, nofootinbib, preprintnumbers, twocolumn, showpacs]{revtex4}

\usepackage{graphicx}
\usepackage{amsmath}
\usepackage{multirow}
\def\fsl#1{\setbox0=\hbox{$#1$}                 
   \dimen0=\wd0                                 
   \setbox1=\hbox{/} \dimen1=\wd1               
   \ifdim\dimen0>\dimen1                        
      \rlap{\hbox to \dimen0{\hfil/\hfil}}      
      #1                                        
   \else                                        
      \rlap{\hbox to \dimen1{\hfil$#1$\hfil}}   
      /                                         
   \fi}                                         %

\newcommand{\VEV}[1]{\langle #1 \rangle}

\begin{document}
\title{Collective excitations, instabilities, and
ground state in dense quark matter}
\date{\today}
\preprint{UWO-TH-06/2}

\author{E.V. Gorbar}
  \email{gorbar@bitp.kiev.ua}
  \altaffiliation[Present address:]{
       Bogolyubov Institute for Theoretical Physics,
       03143, Kiev, Ukraine}
\author{Michio Hashimoto}
  \email{mhashimo@uwo.ca}
\author{V.A. Miransky}
  \email{vmiransk@uwo.ca}
   \altaffiliation[On leave from ]{
       Bogolyubov Institute for Theoretical Physics,
       03143, Kiev, Ukraine}
\affiliation{
Department of Applied Mathematics, University of Western
Ontario, London, Ontario N6A 5B7, Canada}
\author{I.A.~Shovkovy}
\email{shovkovy@th.physik.uni-frankfurt.de}
   \altaffiliation[On leave from ]{
       Bogolyubov Institute for Theoretical Physics,
       03143, Kiev, Ukraine}
\affiliation{Frankfurt Institute for Advanced Studies,
Johann Wolfgang Goethe--Universit\"at,
D-60438 Frankfurt am Main, Germany}
\date{\today}

\begin{abstract}
We study the spectrum of light plasmons in the (gapped and gapless)
two-flavor color superconducting phases and its connection with
the chromomagnetic instabilities and the structure of the
ground state. 
It is revealed that the chromomagnetic instabilities 
in the 4-7th and 8th gluonic channels correspond to two very
different plasmon spectra. These spectra
lead us to the unequivocal conclusion about the existence of 
gluonic condensates 
(some of which can be spatially inhomogeneous)
in the ground state.
We also argue that 
spatially inhomogeneous gluonic condensates 
should exist in
the three-flavor quark matter with the values of the mass of strange
quark corresponding to the gapless color-flavor locked state.

\end{abstract}


\maketitle

It is natural to expect that
cold quark matter
may exist in the interior of compact stars. This fact motivated
intensive studies of this system which
revealed its remarkably rich phase structure
consisting of many different phases (for a review, see Ref. 
\cite{review}).
The question which phase is picked up by nature is still open.

This problem is intimately connected with another one: Recently, it has
been revealed that the $\beta$-equilibrated
gapped (2SC) and gapless (g2SC)
two-flavor color superconducting phases suffer from a chromomagnetic
instability connected with the presence of imaginary (tachyonic)
Meissner masses of gluons~\cite{Huang:2004bg}.
Later a chromomagnetic instability has been also found in the 
gapless three-flavor quark matter \cite{CFL,F}. 

In this Rapid Communication, we will address the problem of the origin of
the chromomagnetic instability and the structure of the ground
state
in cold dense quark matter. The basic
idea underlying the present analysis is the following: While
for calculating (screening) Meissner masses of gluons, it is sufficient
to study the gluon polarization operator only at one point
[$(p_0, \vec{p}) = (0, \vec{p} \to 0)$] in momentum space, we will 
extend this analysis to nonzero energy and momenta and  
study the spectrum of light plasmons. As will be shown, the plasmon 
spectrum
yields an important information about the structure of the
genuine ground state in this system.
Although here only the
two-flavor quark matter will be considered, as 
will be argued below, the present analysis
should be relevant also for the three-flavor case.

Recall that in the two-flavor case, the manifestations of the
chromomagnetic instability are quite different in the regimes
with $\delta\mu < \Delta < \sqrt{2}\delta\mu$ and
$\Delta < \delta\mu$ \cite{Huang:2004bg}, where $\delta\mu$ yields
a mismatch between chemical
potentials for the up and down quarks, and $\Delta$ is
a diquark gap. The (strong coupling) regime with
$\delta\mu < \Delta$
corresponds to the 2SC solution, and the (intermediate coupling)
regime with $\Delta < \delta\mu$ corresponds to the gapless
g2SC one \cite{Shovkovy:2003uu}. While in the g2SC solution both
the 4-7th gluons and the 8th one have tachyonic Meissner masses,
in the 2SC solution, with $\delta\mu < \Delta < \sqrt{2}\delta\mu$, only
the Meissner masses of the 4-7th gluons are tachyonic.

Our analysis reveals that these two chromomagnetic instabilities
correspond to very different spectra of excitations in
the 4-7th and 8th channels. In the 4-7th channels, the
chromomagnetic instability reflects the typical Bose-Einstein
condensation phenomenon: While at subcritical values
of $\delta\mu < \Delta/\sqrt{2}$, there are light plasmons
with the gap (mass) squared 
$0 < {\cal M}^2 \lesssim \Delta^2 \ll \mu^{2}$
($\mu$ is the quark chemical potential), at
supercritical values $\delta\mu > \Delta/\sqrt{2}$, plasmons
become tachyons with ${\cal M}^2 < 0$. 
This picture corresponds
to the conventional continuous (second order) phase transition
and leads us to
the unequivocal conclusion about the existence of gluonic condensates 
in the ground state in the two-flavor case,
like those revealed and described recently in Ref. \cite{gluonic}.

The spectrum of plasmons in the 8th channel is quite different.
There are no light plasmons at all in the 2SC phase    
in that channel. On the other hand, in the g2SC phase, there
is a {\it gapless} tachyonic plasmon with the dispersion relation
$p_{0}^2 = v^2 |\vec{p}|^2$ for small momenta,
where the velocity squared $v^2$ 
is negative.
This collective excitation occurs because
of the existence of gapless modes in the g2SC phase.
The wrong sign of the velocity $v^2$ implies a wrong sign
for the derivative term 
$\partial_i\vec{A}^{\,(8)}\partial_i\vec{A}^{\,(8)}$
in the effective action for gluons.
The latter indicates on a possibility of the
existence of a spatially {\it inhomogeneous} 
gluon condensate $\VEV{\vec{A}^{\,(8)}(\vec{x})}$ in the 
genuine ground state for the values $\delta\mu > \Delta$. 
As we will discuss below,
similar gapless tachyonic plasmons with the quantum
numbers of
$A^{(1)}, A^{(2)}$ gluons and
a linear combination of
diagonal $A^{(3)}, A^{(8)}$ gluons and photon $A^{(\gamma)}$
should exist in the gapless color-flavor locked
(gCFL) phase \cite{Alford:2003fq} in the three-flavor case,
thus signalizing the existence of a spatially inhomogeneous
gluon condensates in the ground state in that phase.
Note that
because the single plane-wave Larkin-Ovchinnikov-Fulde-Ferrell
(LOFF) state \cite{LOFF1,LOFF2} can be characterized by
a homogeneous condensate $\VEV{\vec{A}^{\,(8)}}$ \cite{gluonic}, 
the ground state with spatially inhomogeneous gluon condensates
is an essentially  more complicated medium than
the single plane-wave LOFF one \cite{Gorbar:2005tx}.  
On the other hand, the dynamics with 
a spatially inhomogeneous
gluon condensate $\VEV{\vec{A}^{\,(8)}(\vec{x})}$
is not necessarily inconsistent with
the multiple plane-wave LOFF state \cite{Bowers:2002xr}.
  
Let us consider the polarization tensor for the 4-7th gluons
in the dense two-flavor quark matter in $\beta$-equilibrium.
It is convenient to introduce the fields
$\phi_\mu^+ \equiv \frac{1}{\sqrt{2}}(A_\mu^{(4)}+iA_\mu^{(5)},\,\,
A_\mu^{(6)}+iA_\mu^{(7)})$ and
$\phi_\mu^- \equiv \frac{1}{\sqrt{2}}(A_\mu^{(4)}-iA_\mu^{(5)},\,\,
A_\mu^{(6)}-iA_\mu^{(7)})$, which are color
doublets of the gauge group $SU(2)_c$ in the 2SC/g2SC phase.  
The polarization tensors ${\Pi}_{+}^{\mu\nu}(p_0,\vec p)$ and
${\Pi}_{-}^{\mu\nu}(p_0,\vec p)$ for these fields were
studied in Ref.~\cite{Huang:2004bg} and we will use those
results.  

These polarization tensors are decomposed as
\begin{eqnarray}
\lefteqn{ \hspace*{-0.5cm}
{\Pi}_{\pm}^{\mu\nu}(p_0,\vec p) \equiv
} \nonumber \\ &&
(g^{\mu\nu}-u^\mu u^\nu + 
\frac{{\mathbf p}^\mu {\mathbf p}^\nu }{p^2}) H_{\pm}
+ u^\mu u^\nu K_{\pm} \nonumber \\ &&
- \frac{{\mathbf p}^\mu {\mathbf p}^\mu }{p^2}L_{\pm}
+\left(u^\mu\frac{{\mathbf p}^\nu}{p}+u^\nu\frac{{\mathbf p}^\mu}{p}\right)
  M_{\pm},
\label{decompose}
\end{eqnarray}
where
$p \equiv |\vec p|,\, {\mathbf p}^\mu \equiv (0,\vec p)$,\, and
$u^\mu \equiv (1, 0, 0, 0)$.
The dispersion relations for plasmons are given by \cite{Gusynin:2001tt}
\begin{eqnarray}
\mbox{(magnetic mode):} &&  \qquad p_0^2 - p^2 + H_\pm =0, 
\label{dmag}\\[3mm]
\mbox{(electric mode):~~} && \nonumber \\ && \hspace*{-2.5cm}
  p_0^2 K_{\pm} - p^2 L_{\pm} - 2 p_0 p M_{\pm} + 
  K_{\pm} L_{\pm} + M_{\pm}^2 =0.
\label{delc}
\end{eqnarray}
Note that there are two magnetic modes corresponding to two 
transverse components of plasmons, and one electric mode
corresponding to their longitudinal component. 

We are interested in deriving the gap (mass) spectrum 
${\cal M}_\pm$ of light plasmon excitations, i.e., with
gaps $|{\cal M}_\pm|^2\ll \mu^2$. For this purpose,
we consider the long wavelength limit $|\vec p| \to 0$ with
$p_0$ being finite. 
Because the
color neutrality condition yields
$\mu_8 \sim {\cal O}\left(\frac{\Delta^2}{\mu}\right)$
\cite{Gerhold:2003js}, where $\mu_8$ is the color 
chemical potential,
the approximation with $\mu_8 = 0$
is well justified and will be used here
(the role of a nonzero $\mu_8$ will be clarified below). 
It is easy to check that the following relations are valid 
in this limit and in this approximation:
$L_{\pm}(p_0,|\vec p| \to 0) = H_{\pm}(p_0,|\vec p| \to 0),\,
M_{\pm}(p_0,|\vec p| \to 0) = 0$, and
$K_{\pm}(p_0,|\vec p| \to 0) \neq 0$. Besides that, the
functions $H_{\pm}, L_{\pm}$, and $K_{\pm}$ are of order
$\mu^2$. Therefore,
in the hard dense loop (HDL) approximation that we utilize, 
both dispersion
relations (\ref{dmag}) and (\ref{delc}) for $|\vec p| \to 0$
are reduced to the
equation $H_{\pm}(p_0,|\vec p| \to 0) = 0$. In particular,
the gaps for magnetic and electric modes coincide.

For $\mu_8 = 0$,
the calculations of the function
$H_{\pm}(p_0, 0)$ become
straightforward and we obtain:
\begin{eqnarray}
\lefteqn{\hspace*{-0.3cm}
  H_+(p_0, 0) = H_-(p_0, 0) =
} \nonumber \\ && \hspace*{-0.3cm}
 -\frac{g^2\bar{\mu}^2}{12\pi^2}\Bigg[\,
  4 + \frac{\Delta^2}{p_0^2}
   \ln\left(\,\left(1-\frac{p_0^2}{\Delta^2}\right)^2
               -4\frac{\delta\mu^2}{\Delta^2}
                 \frac{p_0^2}{\Delta^2}\,\right)
\nonumber \\
&& 
+\theta(\delta\mu-\Delta)
 \frac{\Delta^2}{p_0^2}
   \ln\frac{\Delta^4-p_0^2(\delta\mu-\sqrt{\delta\mu^2-\Delta^2})^2}
           {\Delta^4-p_0^2(\delta\mu+\sqrt{\delta\mu^2-\Delta^2})^2}\,\Bigg],
\label{w}
\end{eqnarray}
where $\bar{\mu} \equiv \mu - \delta\mu/3$ and $g$ is the
QCD coupling constant.

\begin{figure}[tbp]
   \begin{center}
     \resizebox{0.45\textwidth}{!}{\includegraphics{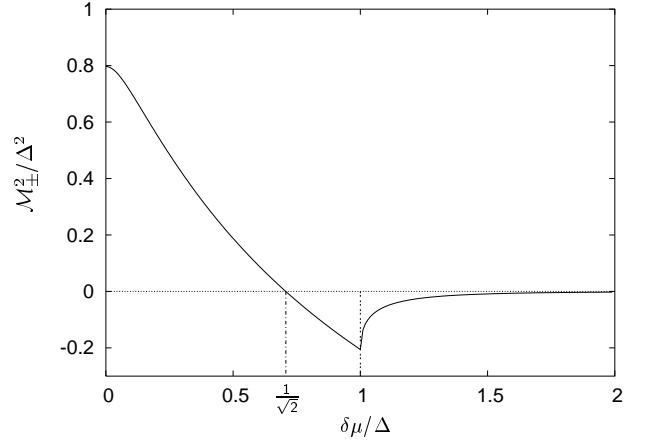}}
   \end{center}
\caption{The plasmon gap squared for 4-7th gluons (solid curve).
\label{fig1}}
\end{figure}

\begin{figure}[tbp]
   \begin{center}
     \resizebox{0.45\textwidth}{!}{\includegraphics{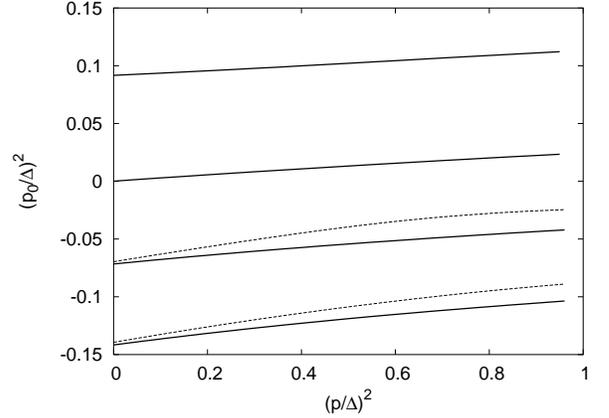}}
   \end{center}
\caption{The dispersion relations for 4-7th magnetic plasmons. The bold 
solid and dashed lines correspond to plasmons in the 2SC and g2SC states, 
respectively. The bold solid lines are for
$\delta\mu/\Delta = 0.6, 1/\sqrt{2}$, 0.8, and 0.9, from top to bottom.
The dashed lines are for
$\delta\mu/\Delta =$ 1.065 and 1.009, from top to bottom.
\label{fig2}}
\end{figure}

For a general
value of $p_0$, the equation $H_{\pm}(p_0, 0) = 0$ 
can be solved only numerically.
However, near the critical point $\delta\mu_{\rm cr} = \Delta/\sqrt{2}$,
we can expand Eq.~(\ref{w}) with respect to $p_0$ and obtain
the analytic solution: 
\begin{equation}
  {\cal M}_\pm \simeq \Delta
  \sqrt{\frac{2}{7}\left(1-2\frac{\delta\mu^2}{\Delta^2}\right)}\;\;\; .
\end{equation}
Only when the mismatch $\delta\mu$ is less than the critical value
$\delta\mu_{\rm cr} = \Delta/\sqrt{2}$, the plasmon gap is real.
For $\delta\mu > \delta\mu_{\rm cr}$, the plasmon gap becomes 
purely imaginary (tachyonic), thus signalizing a Bose-Einstein (BE) 
instability leading to a gluon (plasmon) condensation, as that described
in Ref. \cite{gluonic}.
We also analyzed the case with nonzero $\mu_8$. Its main effect
is in splitting the ${\cal M}_+$ and ${\cal M}_-$ gaps. 

It is noticeable that the BE instability occurs both for 
the magnetic and electric modes. Recall that unlike the Meissner mass,
the (screening) Debye mass for the electric mode remains real for
all values of $\delta\mu$ both in the 2SC and g2SC phases \cite{Huang:2004bg}. 
Therefore the BE instability, connected with the spectrum of plasmons, 
essentially differs from the chromomagnetic instability in this respect.

As was mentioned above, in this approximation,
the values of the gaps for the
magnetic and electric modes are equal. 
As a function of $\delta\mu/\Delta$, their 
gap is shown in
Fig. \ref{fig1}. 
For nonzero momenta, 
the dispersion relations for the magnetic and electric 
modes are different. In the HDL
approximation, we studied these dispersion relations
numerically for all $0 < p^2 \ll \mu^2$. Due to limited space,
here we will describe the results only for the magnetic modes.
Their dispersion relations are plotted in
Fig. \ref{fig2} for several fixed values of $\delta\mu/\Delta$
both in the 2SC and g2SC phases. 
As one can see, the dispersion relations
in the 2SC phase (solid bold lines)
have a qualitative form $p_0^2=m^2+v^2p^2$ with $m^2>0$ for 
$\delta\mu/\Delta<1/\sqrt{2}$ and 
$m^2<0$ for $\delta\mu/\Delta>1/\sqrt{2}$.
The velocity parameter $v$ is real and less than 1. 
The form of dispersion relations in the g2SC phase (dashed
lines) is only slightly different from those in the 
2SC one.
Note that to make the comparison more convenient, the gaps
corresponding to the two dashed lines were chosen to coincide with the 
gaps of the two lower bold solid lines. 

\begin{figure}[tbp]
   \begin{center}
     \resizebox{0.42\textwidth}{!}{\includegraphics{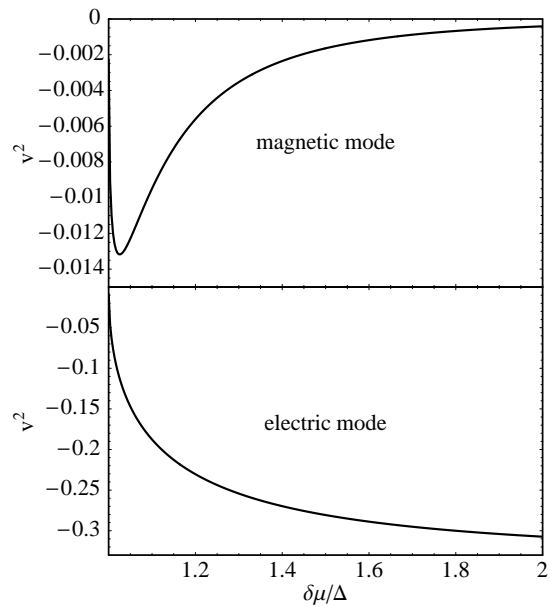}}
   \end{center}
\caption{Velocities squared
for gapless magnetic and electric tachyonic plasmons with
quantum numbers of the 8th gluon. Near the critical point 
$\delta\mu = \Delta + 0$, both velocities have the form
$v^2 \sim -\sqrt{1 - (\Delta/\delta\mu)^2}$. 
\label{fig3}}
\end{figure}

\begin{figure}[tbp]
   \begin{center}
     \resizebox{0.425\textwidth}{!}{\includegraphics{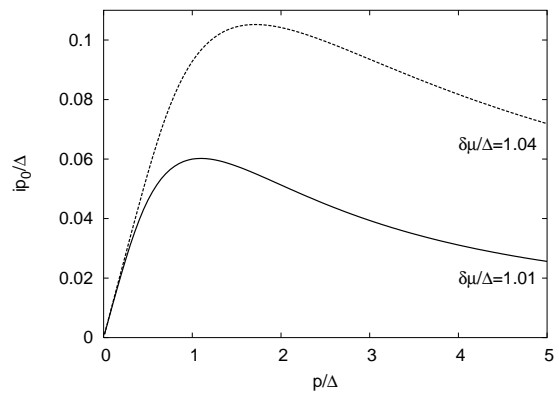}}
   \end{center}
\caption{Dispersion relations
for gapless magnetic tachyonic plasmons with
quantum numbers of the 8th gluon.
\label{fig4}}
\end{figure}

Let us turn to the spectrum of light plasmons in the 8th channel.
The polarization tensor $\Pi_{88}^{\mu\nu}(p_0,\vec p)$
for the $A^{(8)}$ gluon 
can be decomposed as in Eq. (\ref{decompose}) with the
functions $H_{\pm}, K_{\pm}, L_{\pm}$, and $M_{\pm}$ replaced
by $H_{88}, K_{88}, L_{88}$, and $M_{88}$.
In the HDL approximation, 
the dispersion relations for plasmons in this channel
were found numerically.  
The results are the following.
There are no light plasmons in the 8th channel at all in the 2SC phase 
(with $\delta\mu < \Delta$). On the other hand, in the g2SC phase 
(with $\delta\mu > \Delta$), there exist both magnetic and electric
{\it gapless} tachyonic plasmons 
with the dispersion relation
$p_{0}^2 = v^{2} p^2$ as $p \to 0$,
where the velocity squared $v^2$
is negative. These collective excitations occur because
of the existence of gapless modes in the g2SC phase. We
emphasize that there are no light gapped plasmons 
(i.e., with nonzero $p_{0}$ as $p \to 0$) in this
channel. 

In Fig. \ref{fig3}, we depict the velocities squared 
of the gapless magnetic and electric tachyonic plasmons
as functions of $\delta\mu/\Delta$. Like the BE instability
in the 4-7th channels, the gapless tachyon instability 
occurs both for the magnetic and electric modes and, in this respect,
it differs from the chromomagnetic instability in the 8th channel 
in the g2SC phase 
related only to the magnetic mode \cite{Huang:2004bg}.

In Fig.~\ref{fig4},
the dispersion relations for the gapless magnetic plasmon 
are shown for
two different values of $\delta\mu/\Delta$. A notable
feature of these dispersion curves is that the maximum occurs at 
the momentum $p$ of order $\delta\mu\sim\Delta$. A natural
interpretation of this fact would be that this value of
$p$ determines the characteristic scale of an inhomogeneous
gluonic condensate in the genuine ground state.

Because the Meissner mass squared is negative in the
8th channel in the g2SC phase, the existence of the
gapless tachyonic plasmon seems to be counter-intuitive.
What is the mathematical origin of its appearance? The
answer to this question is connected with a peculiar structure
of the function $H_{88}(p_0,p)$. In the kinematic region 
of interest, with $|p_0|, |p|\ll \Delta$, the leading terms in this 
function have the following form:
\begin{eqnarray}
\lefteqn{
H_{88}(p_0,p) =
-\frac{g^2}{3\pi^2}\bar{\mu}^2 \Bigg[\,\frac{4}{3}
      +\left(1-\frac{p_0^2}{p^2}\right)Q\left(\frac{p_0}{p}\right)
} \nonumber \\
&& 
+\frac{1}{2}\theta(\delta\mu-\Delta)
  \frac{\delta\mu}{\sqrt{\delta\mu^2-\Delta^2}}\Bigg\{\,\frac{1}{3}
\nonumber\\
&& \quad
+\left(\,1-\frac{p_0^2}{p^2(1-\frac{\Delta^2}{\delta\mu^2})}\,\right)
       Q\left(\frac{p_0}{p\sqrt{1-\frac{\Delta^2}{\delta\mu^2}}}\right)
       \Bigg\}\Bigg],
\label{H_8}
\end{eqnarray}
where $Q(x)$ is
\begin{eqnarray}
  Q(x) &\equiv& -\frac{1}{2}\int_0^1 d\xi
  \left(\,\frac{\xi}{\xi+x-i\epsilon}
         +\frac{\xi}{\xi-x-i\epsilon}\,\right) 
\nonumber\\
&=& \frac{x}{2} \ln \left|\frac{x+1}{x-1}\right| - 1
  - i \frac{\pi}{2}|x|\theta(1-x^2)\,\,\,(\mbox{for real } x), 
\nonumber\\
&=& y \arctan \frac{1}{y}-1\,\,\, (\mbox{for imaginary } x \equiv iy)
\end{eqnarray}
(subleading terms, such as those of order $\bar\mu^2(p_0/\Delta)^2$,
$\bar\mu^2(p/\Delta)^2$, etc., were omitted here).
Note that the function $H_{88}$ depends on the ratio $p_0/p$.
This fact implies that the point $(p_0, p) = (0, 0)$ is singular:
At $(p_0, p) = (0, 0)$ this function is multivalued and its value depends on
the ratio $p_0/p$ as $p_0, p \to 0$. Therefore this singularity
has many ``faces'' and its manifestations 
are different in different dynamical regimes. While
the screening Meissner mass, related to a static regime
with $p_0/p = 0$, is negative in the g2SC phase, there also exists the
gapless tachyon with the ratio $p_0/p = v$ being purely
imaginary. The physics underlying this behavior of the
function $H_{88}$ is of course
connected with the presence of gapless fermions in the spectrum.

Let us turn to a more detailed discussion
of the results obtained above.
The spectrum of light plasmons in
the 4-7th channels leads us to the conclusion
about the existence of gluonic condensates in
the ground state in the two-flavor case. These condensates 
correspond to the conventional Bose-Einstein 
condensation phenomenon which can
be described in the framework of the Ginzburg-Landau (GL)
approach \cite{gluonic}.

While the instability in the 4-7th channels is
connected with a wrong sign of the gap squared ${\cal M}_{\pm}^2$
and, therefore, with the potential part of the effective
action, the origin of the instability in the 8th channel is
quite different. Indeed, this instability 
is connected with a wrong sign of a velocity
squared $v^2$ of a gapless tachyonic plasmon, 
i.e., with the wrong sign of
the term
$\partial_i\vec{A}^{\,(8)}\partial_i\vec{A}^{\,(8)}$ in
the derivative part of
the effective action. This point is a signal of 
the existence of a spatially inhomogeneous gluon condensate
in the dynamics with $\Delta < \delta\mu$ connected with
the g2SC state. 

A note of caution is in order. As
was shown above, there
exist homogeneous gluonic condensates in the 4-7th
channels 
both for $\delta\mu < \Delta$ and $\delta\mu > \Delta$.
Because
the condensates break spontaneously most of the initial
symmetries in the 2SC state \cite{gluonic}, their dynamics is quite rich
and complex.\footnote{This dynamics
admits a dual, gauge invariant, description as a dynamics
with a condensation of exotic vector mesons \cite{gluonic}.} 
In particular, one cannot 
exclude that these gluonic
condensates themselves could remove the gapless tachyonic plasmon 
in the 8th channel and lead to a consistent spectrum of
excitations in the system.

In fact, the manifestations of a gapless tachyonic instability
could be cleaner in the three-flavor
case. It is known that the lightest value of the
mass $M_s$ of strange quark at which the chromomagnetic instability
occurs is that corresponding to the border between
the CFL and gCFL phases, with $M_{s}^2/\mu \simeq 2\Delta$ \cite{F}.
The instability is generated
in the channels with the quantum numbers  
of $A^{(1)}, A^{(2)}$ gluons 
and a linear combination of
diagonal $A^{(3)}, A^{(8)}$ gluons and photon $A^{(\gamma)}$.
This instability is
similar to the chromomagnetic $A^{(8)}$ instability in 
the g2SC state.  
Therefore, one should expect that 
a gapless tachyonic plasmon
exists in these channels in the gCFL state.
\footnote{In connection with that, we checked that a gapless
tachyonic plasmon exists in a toy model of the gCFL phase
considered in the second paper in Ref. \cite{CFL}.}
It is important that for
this value of
the strange quark mass, no homogeneous gluonic condensates are 
generated (they could occur in the 4-7th channels at larger
values of $M_{s}^2/\mu$ \cite{F}). Therefore, the
phase transition to a spatially inhomogeneous state
in the three-flavor case can be cleaner than in 
the two-flavor one.\footnote{Due
to shortage of space, here we do not consider an influence
of the kaon condensation on instabilities in the gCFL phase
\cite{Schafer:2005ym}.}

The following remark is in order.
Unlike the case of the $A^{(8)}$ gluon in the g2SC phase,
the $A^{(1)},A^{(2)}$ and $A^{(3)},A^{(8)}$  gluons in the gCFL 
phase couple
to gapless quarks with an almost quadratic dispersion
relation \cite{Alford:2003fq}.
However, it is important that
in the vicinity of the transition point to the gapless
phase, from the side of the gapless state, there are hardly any
qualitative differences between these two systems. The reason
is that it is precisely the region where all gapless modes
have an approximately quadratic dispersion law. Because of
that, one should expect that
the transition to the gapless phase
is similar for the cases of 2- and
3-flavor quark matter. Only deeply in the
gapless regime, the analysis in these two cases may develop
qualitative differences.

Although the present analysis does not yield 
a complete answer to the question 
what kind of a transition is connected with
the gapless tachyonic instability, it is clear that it
is not the conventional second order phase transition 
described by the GL effective action. The point is that
a characteristic feature of this gapless tachyon is
that it occurs from ``nothing'': As we emphasized above,
there is no plasmon in the 8th channel in the 2SC state
which is transformed into the gapless tachyonic plasmon in
the g2SC state. 
In other words, there is an abrupt change of the spectrum of
light excitations at the point $\delta\mu = \Delta$ dividing
the 2SC and g2SC phases.

Does it imply that the phase transition connected
with this instability is a first order one? We believe
that this is not necessarily the case: Although the first order
phase transition is a viable option [and it could for example be
similar to that in the model with a p-wave K-meson condensate
\cite{Schafer:2005ym}], there are known examples of a continuous
phase transition with an order parameter going smoothly
to zero at the critical point but with an abrupt change
of the spectrum of light excitations at that same point
\cite{KM}. Such phase transitions are characterized by an essential
singularity in the order parameter at the critical
point. In this regard, it is noticeable that the velocity
squared $v^2$ of gapless tachyonic plasmon goes smoothly
to zero as $\delta\mu/\Delta \to 1 + 0$ (see Fig. \ref{fig2}). On the other
hand, the free energy and Green's functions of gluons
in the 2SC/g2SC state at zero temperature include the
step function $\theta(\delta\mu - \Delta)$ (see Eq. (\ref{H_8})). 
This function introduces
sort of an essential singularity at the border between
the 2SC and g2SC phases and cuts off the dynamics
responsible for light collective excitations in the 
8th channel from the region with $\delta\mu < \Delta$.
This issue deserves further study.

The work of E.V.G, M.H., and V.A.M.
was supported by the Natural Sciences and Engineering Research
Council of Canada.
The work of I.A.S. was supported in part by
the Virtual Institute of the Helmholtz Association
under grant No. VH-VI-041, by the Gesellschaft f\"{u}r
Schwerionenforschung (GSI), and by the Deutsche
Forschungsgemeinschaft (DFG).

\end{document}